\documentclass[twocolumn,groupedaddress,showpacs,preprintnumbers,amsmath,amssymb]{revtex4}

\usepackage{graphicx}
\usepackage{dcolumn}
\usepackage{bm}

\begin{document}
\preprint{Submitted to Journal of Chemical Physics}

\title{Vibron-polaron in $\alpha$-helices. II. Two-vibron bound states}

\author{Cyril Falvo}
\author{Vincent Pouthier}
\email{vincent.pouthier@univ-fcomte.fr}
\affiliation{Laboratoire de Physique Mol\'{e}culaire, UMR CNRS 6624. Facult\'{e} des
Sciences - La Bouloie, \\ Universit\'{e} de Franche-Comt\'{e}, 25030 Besan\c {c}on cedex, 
France.}

\date{\today}

\begin{abstract}

The two-vibron dynamics associated to amide-I vibrations in a 3D $\alpha$-helix is described according to a generalized Davydov model. The helix is modeled by three spines of hydrogen-bonded peptide units linked via covalent bonds.
It is shown that the two-vibron energy spectrum supports both a two-vibron free states continuum and two kinds of bound states, called TVBS-I and TVBS-II, connected to the trapping of two vibrons onto the same amide-I mode and onto two nearest neighbor amide-I modes belonging to the same spine, respectively. 
At low temperature, non vanishing interspine hopping constants yield a three dimensional nature of both TVBS-I and TVBS-II which the wave functions extend over the three spines of the helix. At biological temperature, the pairs are confined in a given spine and exhibit the same features as the bound states described within a one-dimensional model. The interplay between the temperature and the 3D nature of the helix is also responsible for the occurrence of a third bound state called TVBS-III which refers to the trapping of two vibrons onto two different spines.  
The experimental signature of the existence of bound states is discussed through the simulation of their infrared pump-probe spectroscopic response. Finally, the fundamental question of the breather-like behavior of two-vibron bound states is addressed.  

\end{abstract}

\pacs{63.20.Dj,63.22.+m,71.38.Ht,87.10.+e}

\maketitle

\section{Introduction}

In low-dimensional molecular lattices, the nonlinear nature of vibrational excitons (vibrons) plays a key role for energy transfer as well as energy storage in both physical, chemical and biological systems. This feature has become more and more central during the last decade through the concept of discrete breathers \cite{kn:aubry,kn:flach,kn:mackay}. In classical anharmonic lattices, discrete breathers correspond to time-periodic and spatially localized solutions which result from the interplay between the discreteness and the nonlinearity. These solutions do not require integrability for their existence and stability and it has been suggested that they should correspond to quite general and robust solutions. Since discrete breathers yield a local accumulation of the vibrational energy which might be pinned in the lattice or may travel through it, they are expected to be of fundamental importance. 
 
At present, because the occurrence of classical breathers is a relatively well understood phenomena, a great attention is done to characterize their quantum equivalent for which less numerous results are known \cite{kn:fleurov}. In that context, bound states involving two high frequency vibrational excitons, also called two-vibron bound states (TVBS), play a central role. When two vibrons are excited, the vibrational anharmonicity is responsible for the trapping of the two quanta over only a few sites with a resulting energy which is lesser than the energy of two quanta lying far apart. The lateral interaction yields a motion of such a state from one site to another, thus leading to the occurrence of a delocalized wave packet with a well-defined momentum. As a result, TVBS are the first quantum states which experience the nonlinearity and can thus be viewed as the quantum counterpart of breathers or soliton excitations \cite{kn:eilbeck,kn:kimball,kn:bogani,kn:scott}. 

The present paper is thus devoted to the characterization of the TVBS associated to the amide-I vibrations in a 3D $\alpha$-helix. Since the pionneer work of Davydov \cite{kn:davydov,kn:davydov1}, the delocalization of amide-I vibrations yields the occurrence of vibrons which are expected to play a fundamental role for the transport of the energy released by the hydrolysis of adenosine triphophate \cite{kn:scott1,kn:chris}. More precisely, the energy transfer is achieved by small polarons which originate in the strong coupling between the vibrons and the phonons of the helix \cite{kn:brown1,kn:brown2,kn:ivic1,kn:ivic2,kn:ivic3}. A small polaron corresponds to a vibron dressed by a virtual cloud of phonons describing a localized lattice distortion which follows instantaneously the vibron.     

In the paper I of this series, the single-vibron states in a 3D model of an $\alpha$-helix \cite{kn:falvo} have been characterized. It has been shown that the vibron dynamics results from the competition between two kinds of hopping processes. The first kind refers to vibrational transitions between different spines whereas the second kind involves vibron hops along the same spine. This study has revealed that several parameters such as the temperature, the small polaron binding energy and the helix backbone conformation, allow for a transition between two regimes. At low temperature or weak small polaron binding energy, the polaron behaves as an undressed vibron delocalized between the different spines. By contrast, at biological temperature or strong small polaron binding energy, the dressing effect strongly reduces the vibrational exchanges between different spines so that the polaron tends to propagate along a single spine, only. Although the phonon spectrum exhibits both an acoustic branch and an optical branch, it has been shown that the previous features originate in the coupling between the vibrons and the acoustic phonons. 

In recent works \cite{kn:pouthier1,kn:pouthier2}, TVBS have been studied within the 1D model of an $\alpha$-helix in which the vibron dynamics is reduced to that of a single spine of hydrogen-bound peptide unit. In that case, it has been shown that both the intramolecular anharmonicity of each amide-I vibration and the strong vibron-phonon interaction act as nonlinear sources which break the vibron independence and favor the formation of two kinds of bound states. In the harmonic situation, the two bound states appear as combinations of states involving the trapping of the two vibrons onto the same amide-I mode and onto nearest neighbor amide-I modes in a given spine. By contrast, the intramolecular anharmonicity reduces the hybridization between these two kinds of trapping so that the low frequency bound state refers to the trapping of the two vibrons onto the same amide-I mode whereas the high frequency bound state characterizes the trapping onto nearest neighbor amide-I vibrations in a given spine. These results were corroborated by the pump-probe spectroscopy experiment of the N-H mode in a stable $\alpha$-helix which has revealed the two excited state absorption bands connected to the two kinds of bound states \cite{kn:hamm1,kn:hamm2}. 

The present paper is organized as follows. In Sec. II, the model to describe the vibron-phonon dynamics in 3D $\alpha$-helices is introduced. Then, the procedure to remove the intramolecular anharmonicity and to renormalize the vibron-phonon interaction in order to obtain the dressed anharmonic vibron Hamiltonian is summarized. Finally, the number states method used to solve the two-vibron Schrodinger equation is outlined. In Sec. III, a detailed analysis of the two-vibron energy spectrum is performed depending on the values taken by the relevant parameters of the problem. The results are discussed and interpreted in Sec. IV.

\section{Theoretical background}

As pointed out in the introduction, the two-vibron dynamics is strongly sensitive to the nonlinear nature of the vibron-phonon system. In $\alpha$-helices, this nonlinearity originates from both the intramolecular anharmonicity of each amide-I vibration and from the strong vibron-phonon coupling . In the present section, we thus briefly summarize the model described in paper I \cite{kn:falvo} and introduce the anharmonic nature of each amide-I mode according to the 1D model detailed in Ref. \cite{kn:pouthier1}.

\subsection{Helix structure and model Hamiltonian}

Let us consider a sequence of $N$ amino acid units (called residues) regularly distributed along a polypeptide chain and let $n$ the index which labels the position of the $n$th residue. 
In three dimension, the structure of the polypeptide chain is stabilized by the hydrogen bonds between the residues $n$ and $n+3$. The resulting conformation is a 3D $\alpha$-helix in which each residue is related to the next one by a translation of $h=1.5$ \AA\ and by a rotation of $\theta_{0}=100^{o}$ leading to 3.6 residues per turn of helix. The $N$ residues are assumed to behave as point-like entities which the equilibrium positions are located on sites distributed along the helix which the radius
is fixed to $R_{0}=2.8$ \AA\ \cite{kn:davydov1}. The geometry of the helix is thus specified in the (xyz) cartesian frame where the $z$ direction is parallel to the axis of the helix (see Fig. 1a in paper I \cite{kn:falvo}). 

The $n$th site contains a amide-I vibration which behaves as an internal high frequency oscillator described by the standard creation and annihilation vibron operators $b^{+}_{n}$ and $b_{n}$. By introducing the intramolecular anharmonicity of each amide-I mode, the vibron Hamiltonian is written as (within the convention $\hbar$=1)
\begin{eqnarray} 
H_{v}&=& \sum_{n} \omega_{0}b^{\dag}_{n}b_{n}+\gamma_{3}(b^{\dag}_{n}+b_{n})^{3}+\gamma_{4}(b^{\dag}_{n}+b_{n})^{4}+...\nonumber \\
&-&\frac{1}{2} \sum_{n}\sum_{n'} J(n-n') (b^{\dag}_{n}+b_{n})(b^{\dag}_{n'}+b_{n'})
\label{eq:Hv}
\end{eqnarray}
where $\omega_{0}$ stands for the internal frequency of the $n$th amide-I mode and where $J(n-n')$ denotes the lateral hopping constant between the residues $n$ and $n'$. In Eq.(\ref{eq:Hv}), $\gamma_{3}$ and $\gamma_{4}$ represent the cubic and quartic anharmonic parameters of each amide-I mode. 

The amide-I vibrations interact with the phonons of the helix which corresponds to the dynamics of the external motions of the residues. As detailed in paper I \cite{kn:falvo}, this dynamics is expressed in $N$ local frames $(\bm{e}_{r}(n),\bm{e}_{\theta}(n),\bm{e}_{z}(n))$ attached to each residue (see Fig. 1b in paper I \cite{kn:falvo}) so that the displacement of the $n$th residue in the $n$th local frame corresponds to a Bloch wave with wave vector $k$ \cite{kn:chris2}. 
Therefore, the normal mode decomposition is achieved by performing the diagonalization of a $(3 \times 3)$ dynamical matrix $\bm{D}(k)$ for each $k$ value. Such a procedure allows us to define three eigenvalues $\Omega^{2}_{k\sigma}$ and three eigenvectors $\bm{\epsilon}_{k\sigma}$ labeled by the index $\sigma=1,2,3$ associated to the 3D nature of the helix. The two indexes $k$ and $\sigma$ specify a particular phonon mode with energy $\hbar\Omega_{k\sigma}$, quasi-momentum $\hbar k$ and polarization $\bm{\epsilon}_{k\sigma}$. The quantum dynamics of each mode is described by the creation $a^{\dag}_{k\sigma}$ and annihilation $a_{k\sigma}$ operators so that the phonon Hamiltonian is written as (within the convention $\hbar=1$) 
\begin{equation}
H_{p}=\sum_{k\sigma} \Omega_{k\sigma} (a^{\dag}_{k\sigma}a_{k\sigma}+\frac{1}{2})
\label{eq:Hp}
\end{equation}

Finally, according to the Davydov model, the vibron-phonon interaction originates in the modulation of the vibrational frequency of each amide-I vibration by the external motion of the residues. Within the deformation potential approximation, the vibron-phonon coupling Hamiltonian is written in terms of the phonon normal mode coordinates as
\begin{equation}
\Delta H_{vp}=\sum_{k \sigma} (\frac{\Delta_{k\sigma}}{2}e^{-ikn} a^{\dag}_{k\sigma}+\frac{\Delta^{*}_{k\sigma}}{2}e^{ikn} a_{k\sigma})(b^{\dag}_{n}+b_{n})^{2}
\label{eq:Hvp}
\end{equation}
where $\Delta_{k\sigma}$, which accounts for the modulation of the frequency of the $n$th amide-I vibration due to its coupling with the phonon mode specified by the wave vector $k$ and the index $\sigma$, is expressed as 
\begin{equation}
\Delta_{k\sigma}= \sum_{m}  \frac{\chi_{\mid m \mid}\bm{C}_{k}(m)\bm{\epsilon}^{*}_{k\sigma}}
{\sqrt{2MN\hbar \Omega_{k\sigma}}}
\label{eq:DELTA}
\end{equation}
In Eq.(\ref{eq:DELTA}), $\chi_{\mid n-n' \mid}$ denotes the strength of the interaction between the $n$th amide-I mode and the external displacement of the $n'$th residue and $\bm{C}_{k}(m)$ is a vector which account for the helix geometry and which is defined in Eq.(16) of paper I \cite{kn:falvo}.

\subsection{Effective Hamiltonian for the two-vibron dynamics}

The vibron-phonon dynamics is described by the full Hamiltonian $H=H_{v}+H_{p}+\Delta H_{vp}$ which cannot be solved exactly due to the nonlinear sources. Nevertheless, as detailed in Ref. \cite{kn:pouthier1} and outlined in paper I, it can be expressed in an improved way by performing two unitary transformations. First, by following Kimball et al. \cite{kn:kimball}, 
a unitary transformation is applied to remove the intramolecular anharmonicity of each amide-I mode. Then, a modified Lang-Firsov transformation is realized to renormalize a part of the vibron-phonon interaction \cite{kn:lang}. 
Finally, a mean field procedure is applied to separate the vibron degrees of freedom from the phonon coordinates in the residual coupling so that the effective dressed anharmonic vibron Hamiltonian is written as
\begin{eqnarray}
&&\hat{H}_{eff}=\sum_{n} \hat{\omega}_{0}b^{+}_{n}b_{n}-\hat{A}b^{+2}_{n}b_{n}^{2} \nonumber \\
&&-\sum_{nn'} \frac{1}{2} \hat{B}(n-n')b^{+}_{n}b^{+}_{n'}b_{n}b_{n'}  \nonumber \\
&&- \sum_{nn'}J_{1}(n-n')\Phi_{nn'}(N_{n}+N_{n'})b^{+}_{n}b_{n'} \nonumber \\
&&-\sum_{nn'}J_{2}(n-n')\Phi_{nn'}(N_{n}+N_{n'})^{4}b^{+2}_{n}b_{n'}^{2} \nonumber \\
&&- \sum_{nn'}J_{3}(n-n')\Phi_{nn'}(N_{n}+N_{n'})b^{+}_{n}[N_{n}+N_{n'}]b_{n'}
\label{eq:HEFF}
\end{eqnarray}
where the different parameters enter Eq.(\ref{eq:HEFF}) are expressed in terms of the intramolecular anharmonicity 
$A=30\gamma_{3}^{2}/\omega_{0}-6 \gamma_{4}$, the generalized small polaron binding energy $E_{B}(n)=\sum_{k\sigma}\frac{\mid\Delta_{k\sigma}\mid^{2}}{\Omega_{k\sigma}} \cos(kn)$ 
and the anharmonic correction  $\eta=120(\gamma_{3}/\omega_{0})^{2}-12\gamma_{4}/\omega_{0}$ as
\begin{eqnarray}
\hat{\omega}_{0}&=&\omega_{0}-2A-\Sigma_n B(n)/2-(1+4\eta)E_{B}(0) \nonumber \\
\hat{A}&=&A+(1+8\eta)E_{B}(0) \nonumber \\
\hat{B}(n)&=&B(n)+2(1+4\eta)E_{B}(n) \nonumber \\
B(n)&=&144 J(n) (\gamma_{3}/\omega_{0})^{2} \nonumber \\
J_{1}(n)&=&J(n)(1+44(\gamma_{3}/\omega_{0})^{2}-12\gamma_{4}/\omega_{0}) \nonumber \\
J_{2}(n)&=&4 J(n)(\gamma_{3}/\omega_{0})^{2} \nonumber \\
J_{3}(n)&=&J(n)(22(\gamma_{3}/\omega_{0})^{2}-12\gamma_{4}/\omega_{0})
\label{eq:parameter2}
\end{eqnarray}
In Eq.(\ref{eq:HEFF}), $\Phi_{nn'}(X)=\exp(-S(n-n',T)[1+2\eta+2\eta X])$ where $S(n,T)$ is the coupling constant 
defined as ($k_{B}$ denotes the Boltzmann constant)
\begin{equation}
S(n,T)=\sum_{k\sigma} \mid \frac{\Delta_{k\sigma}}{\Omega_{k\sigma}} \mid^2 \coth(\frac{\hbar \Omega_{k\sigma}}{2k_BT})(1-\cos(kn))
\label{eq:S}
\end{equation}

As clearly shown in Eq.(\ref{eq:HEFF}), the nonlinear sources strongly modify the dynamics of the dressed anharmonic vibrons. They are first responsible for a redshift of each amide-I frequency, i.e. $\omega_{0} \rightarrow \hat{\omega_{0}}$. Then, they lead to the occurrence of coupling terms which break the vibron independence. The terms $\hat{A}b^{+2}_{n}b_{n}^{2}$ and $\hat{B}(n-n')b^{+}_{n}b^{+}_{n'}b_{n}b_{n'}$ yield an attractive interaction between two vibrons and favor their trapping around neighboring amide-I sites. Finally, the nonlinearities modify the vibron hopping processes via the dressing effect which reduces the hopping constants connecting two sites $n$ and $n'$ according to the dressing function $\Phi_{nn'}$. Note that the contributions proportional to $J_{2}$ characterize hops in the course of which two vibrons realize simultaneously a transition whereas the terms proportional to $J_{3}$ affect single-vibron hops from states formed by two vibrons located onto the same amide-I vibration. These two latter mechanism are specific of the two-vibron dynamics and do not occur in the single-vibron dynamics. At this step let us mention that the 3D nature of the helix provides a site dependent small polaron binding energy $E_{B}(n)$ as well as a site dependent coupling constant $S(n-n',T)$. As it will be discussed in the following section, these two features discriminate between interspine and intraspine processes for both vibron hops and vibron-vibron interactions. 

To characterize the two-vibron dynamics, the corresponding Schrodinger equation $\hat{H}_{eff}\mid \Psi \rangle = \omega \mid \Psi \rangle$ has to be solved. Since the Hamiltonian $\hat{H}_{eff}$ (Eq.(\ref{eq:HEFF})) conserves the number of vibrons, this can be achieved by using the number states method \cite{kn:scott} which was successfully applied to molecular adsorbates \cite{kn:pouthier01,kn:pouthier02,kn:pouthier03} and 1D model of $\alpha$-helices \cite{kn:pouthier1,kn:pouthier2}. 
Within this method, the two-vibron wave function is expanded as  
\begin{equation}
\mid \Psi  \rangle = \sum_{n_{1},n_{2}\geq n_{1}} \Psi(n_{1},n_{2}) 
\mid n_{1},n_{2} ) 
\label{eq:basis}
\end{equation}
where $\{\mid n_{1},n_{2} ) \}$ denotes a local basis set normalized and symmetrized according to the restricting $n_{2} \geq n_{1}$ and where a particular vector $\mid n_{1},n_{2} )$ characterizes two vibrons located onto the sites $n_{1}$ and $n_{2}$, respectively. This basis set generates the entire two-vibron subspace which the dimension $N(N+1)/2$ represents the number of ways for distributing two indistinguishable quanta onto $N$ sites. 

The Schrodinger equation can be expressed in an improved way by taking advantage of the helix symmetry.
Indeed, as for the phonon dynamics, the two-vibron wave function is invariant under a rotation by an angle of $\theta_0$ around the helix axis followed by a translation along this axis by a distance $h$. As a result, it can be expanded as a Bloch wave as
\begin{equation}
\Psi(n_{1},n_{2}=n_{1}+m)= \frac{1}{\sqrt{N}} \sum_{n_{1}} e^{iK(n_{1}+m/2)}\Psi_{K}(m)
\label{eq:bloch}
\end{equation}
where the total momentum $K$, which takes $N$ values belonging to the first Brillouin zone of the helix, is associated to the motion of the center of mass of the two vibrons. Therefore, the resulting wave function $\Psi_{K}(m)$ refers to the degree of freedom $m$ which characterizes the distance between the two vibrons along the helix. Since the momentum $K$ is a good quantum number, the Hamiltonian $H_{eff}$ appears as block diagonal and the Schrodinger equation can be solved for each $K$ value. This procedure, achieved numerically, allows us to defined a set of two-vibron eigenenergies $\omega_{K\lambda}$ and a set of two-vibron eigenstates $\Psi_{K\lambda}(m)$ which are specified by the index $\lambda=1,...,(N+1)/2$ for each $K$ value.

In the following section, this procedure is illustrated to characterize the two-vibron energy spectrum of a 3D $\alpha$-helix.

\section{Numerical results}

\begin{figure*}
\begin{center}
\begin{minipage}[c]{0.68\linewidth}
\includegraphics{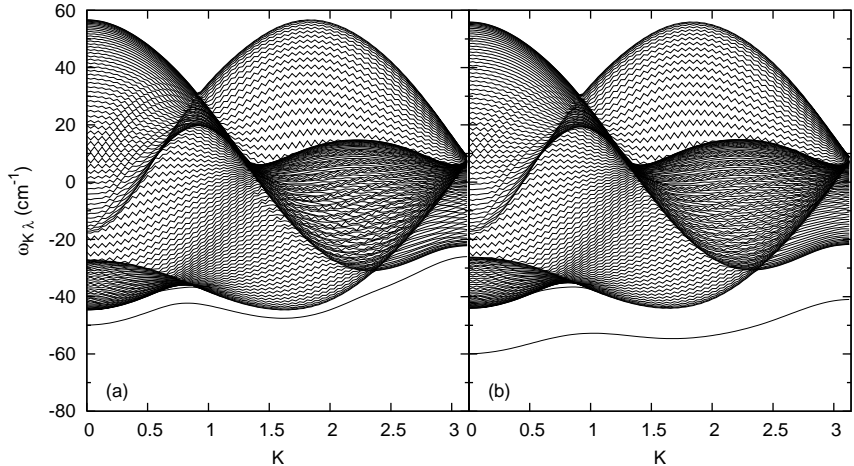}
\end{minipage}\hfill
\begin{minipage}[c]{0.32\linewidth}
\caption{Two-vibron energy spectrum for $T=5$ K, $\chi_{1}=25$ pN, $\chi_{3}=50$ pN and for (a) $A=0$ and (b) $A=8$ cm$^{-1}$. The spectrum is centered onto the corrected frequency $2\hat{\omega}_{0}$ and corresponds to the two-vibron dispersion curves drawn in half of the first Brillouin zone of the helix, i.e. $0<K<\pi$.}
\end{minipage}
\end{center}
\end{figure*}

To describe the vibron dynamics in 3D $\alpha$-helices, the hopping constant along the hydrogen bonds is fixed to $J(3)=7.8$ cm$^{-1}$ \cite{kn:scott1,kn:ivic2} whereas the hopping constants between different spines of hydrogen-bonded peptide units, calculated in Ref. \cite{kn:scott2} are equal to $J(1)=-12.4$ cm$^{-1}$ and $J(2)=3.9$ cm$^{-1}$. The other elements of the hopping matrix, rather small, are listed in the Table I of Ref. \cite{kn:scott2}. The harmonic frequency is fixed to $\omega_{0}=1695$ cm$^{-1}$ and the anharmonic constant is equal to $A=8.0$ cm$^{-1}$ \cite{kn:pouthier1,kn:hamm3,kn:hamm4}. The cubic and quartic anharmonic parameters can be expressed approximately by using the relation $15\gamma_{3}^{2}/\omega_{0} \approx 6 \gamma_{4}\approx A$ \cite{kn:pouthier1}. Note that these parameter values are in a rather good agreement with recent ab initio calculations \cite{kn:muk}.

As discussed in paper I \cite{kn:falvo}, the harmonic dynamics of the phonons is essentially governed by the force constant $K_{\mid n-n'\mid}$ which refers to the second derivative of the pair potential connecting the two residues located onto the sites $n$ and $n'$. The force constant $K_{2}$ is set to zero and we assume $K_{\mid n \mid}=0$ for $n\geq4$. The force constant of the hydrogen bonds is fixed to $K_{3}=15$ Nm$^{-1}$ \cite{kn:scott1,kn:ivic2,kn:pouthier1,kn:pouthier2,kn:hennig} whereas the force constant of the covalent bonds is equal to $K_{1}=60$ Nm$^{-1}$ \cite{kn:hennig}. Finally, the mass $M$ which enters the phonon dynamics has been fixed to 2.0 $10^{-25}$ kg. The strength of the vibron-phonon coupling Eq.(\ref{eq:Hvp}) is given by the parameters $\chi_{\mid n-n'\mid}$ \cite{kn:falvo} which accounts for the modulation of the $n$th amide-I frequency due to the external motion of the $n'$th residue. According to the 1D Davydov model, the parameter $\chi_{3}$ ranges between 35 and 62 pN and we  
choose $\chi_{2}=0$ whereas we treat $\chi_{1}$ as a parameter smaller than $\chi_{3}$ \cite{kn:hennig}. Note that $\chi_{n}=0$ for $n \geq 4$.

The two-vibron energy spectrum is shown in Figs. 1 for $T=5$ K, $\chi_{1}=25$ pN and $\chi_{3}=50$ pN ($E_{B}(0)=8.9$ cm$^{-1}$). The spectrum, centered onto the corrected frequency $2\hat{\omega}_{0}$ (Eq.(\ref{eq:parameter2})), describes the two-vibron dispersion curves drawn in half of the first Brillouin zone of the helix, i.e. $0<K<\pi$.
In the harmonic approximation, i.e. $A=0$ (Fig. 1a), the spectrum exhibits an energy continuum located around $2\hat{\omega}_{0}$ with a bandwidth equal to 101.9 cm$^{-1}$. This continuum contains the states describing two independent vibrons and called two-vibron free states (TVFS) \cite{kn:pouthier1}. For a given $K$ value, a free state refers to two independent vibrons with wave vectors $q_{1}$ and $q_{2}$ and energy $\omega_{q_{1}}$ and $\omega_{q_{2}}$ so that $K=q_{1}+q_{2}$ and $\omega_{K\lambda}=\omega_{q_{1}}+\omega_{q_{2}}$. Below the continuum, the spectrum exhibits two bands connected to two different bound states. The low frequency band, located below the TVFS over the entire Brillouin zone, refers to bound states called TVBS-I. The binding energy of TVBS-I, i.e. the gap between the zero wave vector TVBS-I and the bottom of the TVFS continuum, is equal to 4.8 cm$^{-1}$ whereas its bandwidth is about 24.0 cm$^{-1}$. The second band, which refers to bound states denoted TVBS-II, lies below the continuum in a wave vector window which ranges between $K=0.5$ and $K=0.9$. When the anharmonic parameter is set to $A=8$ cm$^{-1}$ (Fig. 1b), the shape of the TVFS continuum does not exhibit any significant modification. Note that the center of the continuum is redshifted due to the dependence of $2\hat{\omega}_{0}$ (Eq.(\ref{eq:parameter2})) on the anharmonic parameter $A$ (not drawn in the figure). Nevertheless, the TVFS bandwidth is slightly reduced to 100.0 cm$^{-1}$ so that the wave vector window for the occurrence of the TVBS-II opens and ranges between $K=0.4$ and $K=1.0$. By contrast, the anharmonicity strongly modifies the TVBS-I properties. It is responsible for a strong redshift of the band as well as for a decrease of the bandwidth. The TVBS-I binding energy reaches 15.3 cm$^{-1}$ whereas the bandwidth is reduced to 19.0 cm$^{-1}$. 

\begin{figure}
\includegraphics{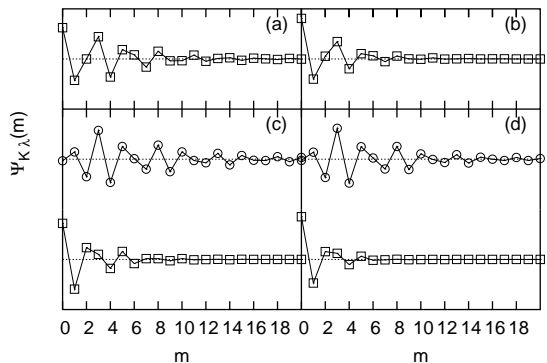}
\caption{TVBS wave function $\Psi_{K\lambda}(m)$ for $T=5$ K, $\chi_{1}=25$ pN, $\chi_{3}=50$ pN. (a) $K=0$ and $A=0$ cm$^{-1}$, (b) $K=0$ and $A=8$ cm$^{-1}$, (c) $K=0.69$ and $A=0$ cm$^{-1}$(d) $K=0.69$ and $A=8$ cm$^{-1}$.The TVBS-I wave function is described by open squares whereas the TVBS-II wave function is represented by open circles.}
\end{figure}
\begin{figure*}
\begin{center}
\begin{minipage}[c]{0.68\linewidth}
\includegraphics{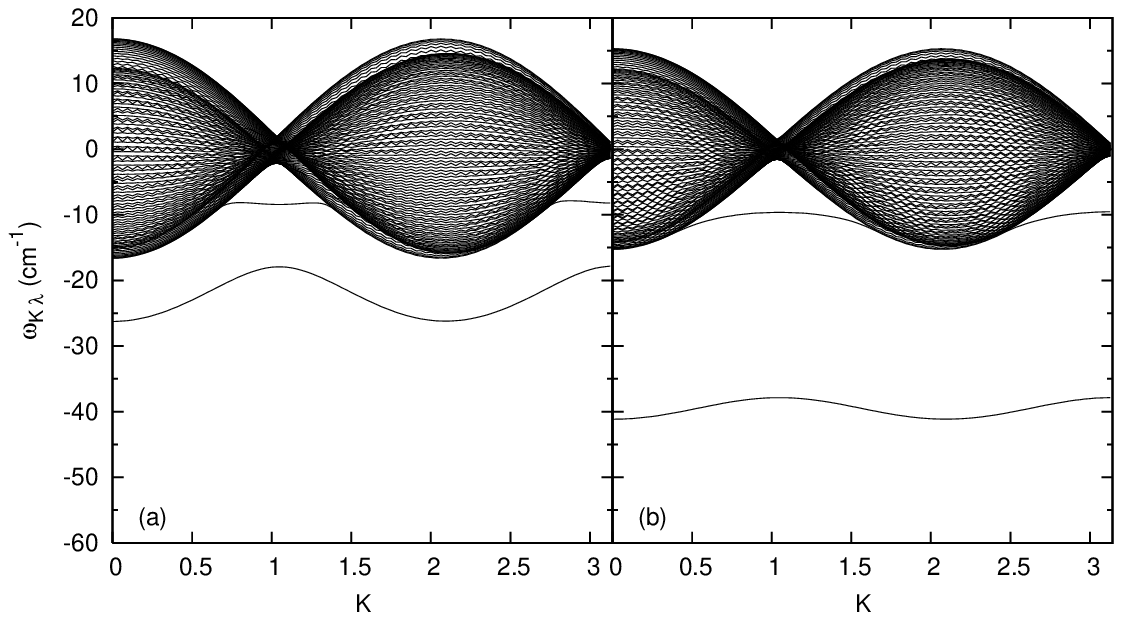}
\end{minipage}\hfill
\begin{minipage}[c]{0.32\linewidth}
\caption{Two-vibron energy spectrum for $T=310$ K, $\chi_{1}=25$ pN, $\chi_{3}=50$ pN and for (a) $A=0$ and (b) $A=8$ cm$^{-1}$. }
\end{minipage}
\end{center}
\end{figure*}
Figs. 2 displays the TVBS wave function $\Psi_{K\lambda}(m)$ corresponding to the situation described in Figs. 1. Figures 2a and 2b show the zero wave vector TVBS-I wave function for $A=0$ and $A=8$ cm$^{-1}$, respectively. The same curves are presented in figures 2c and 2d but for a wave vector equal to $K=0.69$. The TVBS-I wave function is described by open squares whereas the TVBS-II wave function is represented by open circles.  
When $K=0$, the spectrum supports a single bound state, namely TVBS-I (Figs. 1). When $A=0$ (Fig. 2a), the TVBS-I wave function is maximum when $m=0$ and slightly decreases as the separating distance $m$ between the two vibrons increases. It exhibits oscillations and takes significant values for $m=1$ and $m=3$ whereas it almost vanishes when $m=2$. When the anharmonicity is set to $A=8$ cm$^{-1}$ (Fig. 2b), the maximum in $m=0$ is reinforced whereas the extension of the wave function has been reduced. In other words, such a state refers to two vibrons trapped around the same amide-I vibration. However, whatever the anharmonicity, the wave function shows a 3D character so that the bound state extends over the three spines of the helix. 
When $K=0.69$, Figs. 1 clearly shows that the energy spectrum supports both TVBS-I and TVBS-II. The TVBS-I wave function appears strongly localized onto $m=0$ and  exhibits basically the same features as when $K=0$ (Figs. 2c and 2d). By contrast, the TVBS-II wave function vanishes in $m=0$. It is maximum in $m=3$ and decreases as the separating distance between the two vibrons increases or decreases from $m=3$ (Figs. 2c and 2d). Such a state refers to two vibrons trapped around two nearest neighbor amide-I vibrations in a given spine. 
Nevertheless, the TVBS-II wave function shows a 3D character so that the three spines of the helix are involved in the extension of the bound state. 
 
\begin{figure}
\includegraphics{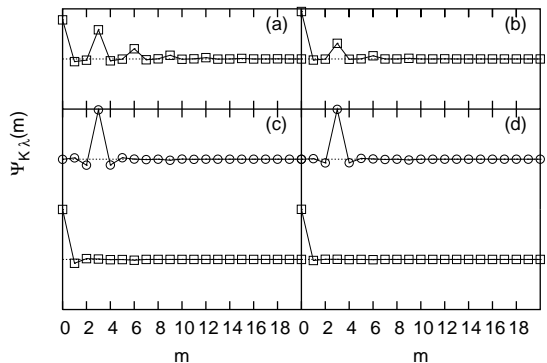}
\caption{TVBS wave function $\Psi_{K\lambda}(m)$ for $T=310$ K, $\chi_{1}=25$ pN, $\chi_{3}=50$ pN. (a) $K=0$ and $A=0$ cm$^{-1}$, (b) $K=0$ and $A=8$ cm$^{-1}$, (c) $K=\pi/3$ and $A=0$ cm$^{-1}$(d) $K=\pi/3$ and $A=8$ cm$^{-1}$.The TVBS-I wave function is described by open squares whereas the TVBS-II wave function is represented by open circles.}
\end{figure} 
\begin{figure*}
\begin{center}
\begin{minipage}[c]{0.68\linewidth}
\includegraphics{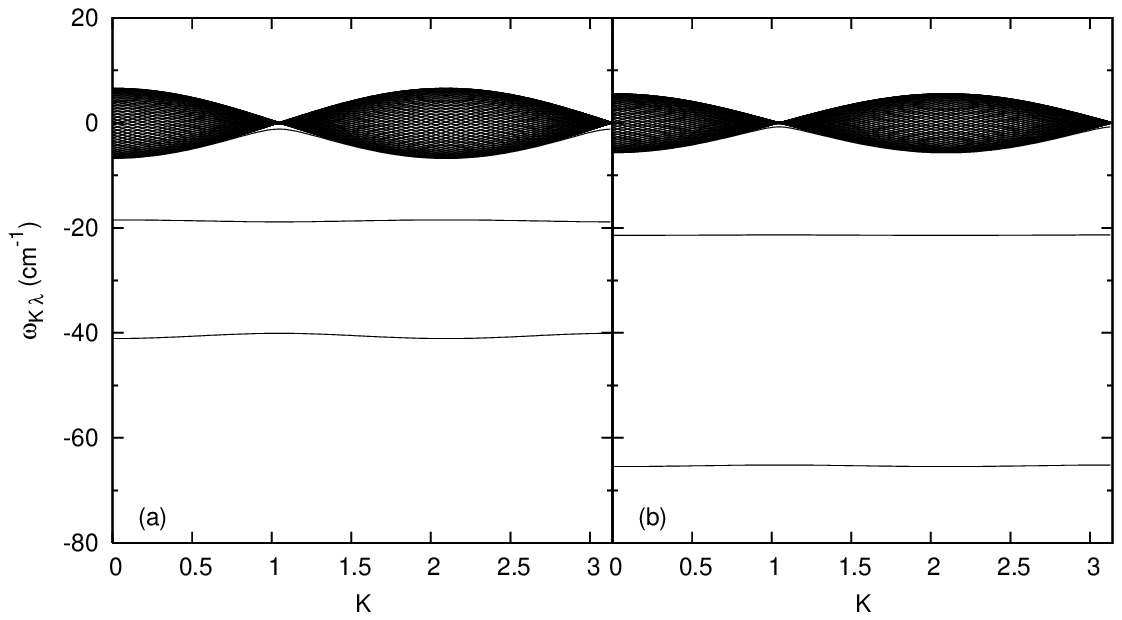}
\end{minipage}\hfill
\begin{minipage}[c]{0.32\linewidth}
\caption{Two-vibron energy spectrum for $T=310$ K, $\chi_{1}=37.5$ pN, $\chi_{3}=75$ pN and for (a) $A=0$ and (b) $A=8$ cm$^{-1}$.}
\end{minipage}
\end{center}
\end{figure*}
\begin{figure}
\includegraphics{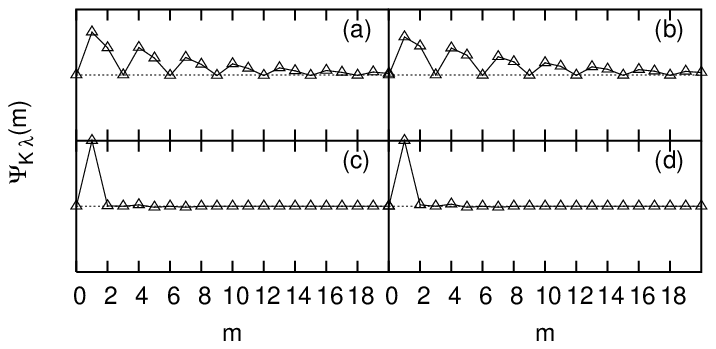}
\caption{TVBS-III wave function $\Psi_{K\lambda}(m)$ for $T=310$ K, $\chi_{1}=37.5$ pN, $\chi_{3}=75$ pN. (a) $K=\pi/4$ and $A=0$ cm$^{-1}$, (b) $K=\pi/4$ and $A=8$ cm$^{-1}$, (c) $K=\pi/3$ and $A=0$ cm$^{-1}$(d) $K=\pi/3$ and $A=8$ cm$^{-1}$.}
\end{figure}
In Figs. 3, the two-vibron energy spectrum is illustrated for $T=310$ K and for the same set of parameters used in Figs. 1. Whatever the anharmonicity, the increase of the temperature strongly modifies the nature of the 
TVFS continuum. It appears symmetrically distributed around $2\hat{\omega_{0}}$ with a bandwidth which has been strongly reduced. It is equal to 33.4 cm$^{-1}$ when $A=0$ (Fig. 3a) whereas it reaches 30.5 cm$^{-1}$ when $A=8$ cm$^{-1}$ (Fig. 3b). As in Figs. 1, the TVBS-I lies below the continuum over the entire Brillouin zone. However, its binding energy increases at biological temperature whereas the corresponding bandwidth decreases. In the harmonic situation, i.e. $A=0$ (Fig. 3a), the TVBS-I binding energy is equal to 9.7 cm$^{-1}$ and the bandwidth is equal to 8.4 cm$^{-1}$. When the anharmonicity is fixed to $A=8$ cm$^{-1}$ (Fig. 3b), the binding energy reaches 25.9 cm$^{-1}$ whereas the bandwidth decreases to 3.6 cm$^{-1}$. As shown in Figs. 3, the spectrum displays now two wave vector windows for the occurrence of the TVBS-II. For $A=0$ (Fig. 3a), a first wave vector window appears in the range $K= 0.7 $ and $K= 1.4 $ and a second window takes place at the end of the first Brillouin zone, i.e. for $K> 2.7$. As previously, as when the anharmonicity is increased, the range of these two windows increases (Fig. 3b) so that the TVBS-II band clearly tends to get out of the continuum.

The corresponding TVBS wave functions are illustrated in Figs. 4 for $A=0$ (Figs. 4a and 4c) and $A=8$ cm$^{-1}$ (Fig. 4b and 4d). When $K=0$ (Figs. 4a and 4b), the spectrum supports the TVBS-I band, only. Whatever the anharmonicity, the TVBS-I wave function is maximum in $m=0$ and takes significant values for $m=3$, $m=6$, $m=9$ ... and so on. When the anharmonicity increases, the maximum in $m=0$ is reinforced and the extension of the wave function is reduced. In other words, at biological temperature, the two vibrons are confined in a given spine and are trapped around the same amide-I mode. When $K=\pi/3$ (Figs. 4c and 4d), Figs. 3 clearly show that the spectrum supports both TVBS-I and TVBS-II. The TVBS-I wave function is strongly localized onto $m=0$ and almost vanishes for the other $m$ values. By contrast, the TVBS-II wave function is clearly localized onto $m=3$ so that it refers to two vibrons strongly trapped onto two nearest neighbors amide-I modes belonging to the same spine. Note that the anharmonicity does not modify significantly the nature of the wave function of both TVBS-I and TVBS-II.    

Finally, the two-vibron energy spectrum is illustrated in Figs. 5 for $T=310$ K$, \chi_{1}=37.5$ pN and $\chi_{3}=75.0$ pN ($E_{B}(0)=20.0$ cm$^{-1}$). In that case, the strong vibron-phonon coupling is responsible for a drastic decrease of the TVFS bandwidth. As in Figs. 3, the anharmonicity enhances the decrease of the bandwidth which varies from 13.3 cm$^{-1}$ for $A=0$ (Fig. 5a) to 11.2 cm$^{-1}$ for $A=8$ cm$^{-1}$ (Fig. 5b). As previously, the spectrum exhibits a TVBS-I band located below the continuum. The anharmonicity reduces the corresponding bandwidth and yields a strong redshift of the band. However, the main difference when compared with the two previous cases is twofold. First, the dressing effect is strong enough to induce the occurrence of the TVBS-II band over the entire Brillouin zone, even when $A=0$ (Fig. 5a). The anharmonicity increases the TVBS-II binding energy and reduces its bandwidth. Then, a third bound state band, called TVBS-III, occurs below the TVFS continuum. 
Although it does not appear clearly on the figure, this band lies below the continuum over the entire Brillouin zone. In a marked contrast with the two previous bound states, the TVBS-III binding energy decreases as when the anharmonicity increases since it varies from 0.08 cm$^{-1}$ when $A=0$ (Fig. 5a) to 0.076 cm$^{-1}$ when $A=8$ cm$^{-1}$ (Fig. 5b). By contrast, the TVBS-III bandwidth decreases with $A$. It is equal to 5.59 cm$^{-1}$ when $A=0$ (Fig. 5a) and reaches 4.94 cm$^{-1}$ when $A=8$ cm$^{-1}$ (Fig. 5b).  
 
In the latter situation corresponding to a strong dressing effect, we have verified that for both the TVBS-I and the TVBS-II, the two vibrons are confined in a given spine. TVBS-I refers to their trapping onto the same amide-I mode ($m=0$) whereas TVBS-II characterizes their trapping onto two nearest neighbor amide-I modes ($m=3$) located in the same spine (the wave functions are not drawn). The behavior of the wave function of the third bound state is illustrated in Figs. 6 for $A=0$ (Figs. 6a and 6c) and $A=8$ cm$^{-1}$ (Figs. 6b and 6d). When $K=\pi/4$ (Figs. 6a and 6b), the TVBS-III wave function exhibits non vanishing values for $m=1$, $2$, $4$, $5$, ... and so on, with a resulting amplitude which decreases as $m$ increases. By contrast, when $K=\pi/3$ (Figs. 6c and 6d), the wave function appears strongly localized onto $m=1$, only. These features indicate that TVBS-III refers to vibrons trapped onto neighboring amide-I modes belonging to two different spines. More precisely, when the TVBS-III binding energy is rather weak, for instance when $K=\pi/4$, the bound state is extended and the two vibrons are trapped over a large distance but belong to two different spines. By contrast, the TVBS-III tends to localized when its binding energy increases so that it refers to vibrons trapped onto two nearest neighbor amide-I modes belonging to two different spines.

\section{Discussion}

To interpret and discuss the previous results, let us first focus our attention onto the influence of the nonlinear sources, i.e. the intramolecular anharmonicity and the strong vibron-phonon coupling. 
As shown in Figs. 1, 3 and 5, the intramolecular anharmonicity is responsible for a decrease of the TVFS bandwidth, typically of about a few wave numbers. This feature originates in the modification of the hopping constants due to both the anharmonicity and the dressing effect. Indeed, a TVFS corresponds to two independent vibrons moving according to the effective hopping constants $J_{eff}(n)=J_{1}(n)exp[-(1+4\eta)S(n,T)]$ which differ from the constants involved in the harmonic approximation. Basically, the intramolecular anharmonicity reinforces the role played by the coupling constant $S(n,T)$ which characterizes the dressing effect ($S(n,T)\rightarrow (1+4\eta)S(n,T)$). Therefore, it enhances the dressing effect and favors a decrease of the effective hopping constants so that anharmonic vibrons are more sensitive to the dressing effect than harmonic vibrons. Note that as when the temperature is increased, the strength of the dressing effect increases which results in a drastic decrease of the TVFS bandwidth. 
As shown in Figs. 1, 3 and 5, the TVBS-I are located below the TVFS continuum over the entire Brillouin zone whereas, for TVBS-II, two situations occur depending on the strength of the vibron-phonon coupling. For a weak coupling, the band disappears inside the continuum when the wave vector lies outside specific windows whereas for a strong coupling, the band is located below the continuum over the entire Brillouin zone. As when increasing the anharmonicity, the TVBS-I band is redshifted and its bandwidth is strongly reduced. In the same way, the anharmonicity modifies the nature of the TVBS-II band. If the band is resonant with the continuum, the anharmonicity tends to open the wave vector windows and when  the TVBS-II band is not resonant with the continuum, the anharmonicity induces a redshift of the band as well as a decrease of its bandwidth. 
Note that these different features generalize the work detailed in Ref. \cite{kn:pouthier1} devoted to the two-vibron dynamics in one-dimensional model of helix. We thus can conclude that the occurrence of both TVBS-I and TVBS-II originates in the interplay between the intramolecular anharmonicity $A$ and the site dependent small polaron binding energy $E_{B}(n)$. More precisely, in a 3D helix, it is straightforward to show that the small polaron binding energy takes significant values for $n=0$ and $n=3$ and satisfies $E_{B}(0) \approx 2E_{B}(3)$ in a perfect agreement with the results obtained in Ref. \cite{kn:pouthier1}. Therefore, $A$ and $E_{B}(0)$ are the key parameters which favors the occurrence of the TVBS-I through the terms $\hat{A}b_{n}^{\dag}b_{n}^{\dag}b_{n}b_{n}$ in Eq.(\ref{eq:HEFF}) which characterize an attractive interaction between two vibrons located onto the same amide-I mode. 
By contrast, the occurrence of TVBS-II is essentially controlled by the parameter $E_{B}(3)$ which, as shown in Eq.(\ref{eq:HEFF}), yields an attractive interaction between two vibrons located onto two nearest neighbor amide-I modes belonging to the same spine. To obtain a TVBS-II band below the TVFS continuum over the entire Brillouin zone, $E_{B}(3)$ must exceed a critical value typically of about $2J_{eff}(3)$. As a consequence, TVBS-I refers to the trapping of the two vibrons onto the same amide-I mode whereas TVBS-II characterizes two vibrons trapped onto two nearest neighbor amide-I modes located in the same spine.

However, the main results of the present study concerns the influence of the temperature on the nature of the two-vibron states in a real helix. 

As detailed in paper I \cite{kn:falvo}, the single-vibron dynamics in a 3D $\alpha$-helix is characterized by two kinds of hopping processes. The first kind refers to vibrational transition between different spines of hydrogen-bonded peptide units whereas the second kind involves vibron hops along the same spine. In that context, the nature of the vibron dynamics essentially results from the competition between these two kinds of processes. This previous study has revealed that several parameters such as the temperature, the small polaron binding energy and the helix backbone conformation, allow for a transition between two regimes. At low temperature the small polaron behaves like an undressed vibron so that the 3D nature of the helix plays a crucial role and the vibron is delocalized between the different spines. Within this undressed limit, vibron transitions between nearest neighbor and third nearest neighbor sites represent the dominant hopping mechanism since $J(1)=-12.4$ cm$^{-1}$ and $J(3)=7.8$ cm$^{-1}$ whereas $J(2)=3.9$ cm$^{-1}$. By contrast, at biological temperature or strong small polaron binding energy, the dressing effect strongly reduces the vibrational exchanges between different spines so that the polaron tends to propagate along a single spine, only. Indeed, the dressing for vibrons hops between different spines is more efficient than the dressing which affects vibron transitions along the same spine so that $J_{eff}(3)$ becomes the dominant hopping constant. 

The singular behavior of the single vibron dynamics is responsible for the occurrence of a strong temperature dependence on the two-vibron states. 
To illustrate this feature, let us first consider the temperature dependence of the TVFS continuum. Indeed, Figs. 1 clearly show that at low temperature the continuum exhibits a large bandwidth characterized by a complicated structure whereas at biological temperature (Figs. 3 and 5), it appears symmetrically distributed around the central frequency $2\hat{\omega_{0}}$. In that context, it is straightforward to show that the continuum is perfectly reproduced from the knowledge of the single vibron energy spectrum and that the TVFS
bandwidth is twice the single vibron bandwidth. Therefore, since single vibron states are delocalized over the three spines at low temperature, the different hopping constants contribute to the bandwidth. For instance, when $T=5$ K, $\chi_{1}=25$ pN, $\chi_{3}=50$ pN and $A=8$ cm$^{-1}$, the single vibron bandwidth is about 50 cm$^{-1}$ and yields a TVFS bandwidth equal to 100.0 cm$^{-1}$ in a perfect agreement with the results displayed in Figs. 1. Note that in that case the bandwidth mainly depends on both $J_{eff}(1)$ and $J_{eff}(3)$. By contrast, at biological temperature, the continuum displayed in Figs. 3 and 5 mainly refers to two independent vibrons confined in a given spine. When long range hopping constants are disregarded, the single vibron bandwidth is equal to $4J_{eff}(3)$. Therefore, when $T=310$ K, $\chi_{1}=25$ pN and $\chi_{3}=50$ pN, this value leads to a TVFS bandwidth equal to 29.12 cm$^{-1}$ when $A=8$ cm$^{-1}$, in a perfect agreement with the value characterizing the real continuum (see Figs. 3). 
Note that the shape of the continuum results form the unfolding of the two-vibron dispersion curves of the 1D model since the Brillouin zone of the helix is three time greater than the Brillouin zone of the corresponding 1D model. Moreover, long range effects in the hopping constants, which have been disregarded in Ref.\cite{kn:pouthier1}, lead to non vanishing values of the continuum for both $K=\pi/3$ and $K=\pi$. 

The influence of the temperature on the two-vibron bound states is twofold. First, it strongly modifies the nature of both TVBS-I and TVBS-II. Then, it allows for the occurrence of a third bound state, i.e. TVBS-III, which has not been obtained within our previous one-dimensional model \cite{kn:pouthier1}.

At low temperature (see Figs. 2), non vanishing interspine hopping constants yield a three dimensional nature of both TVBS-I and TVBS-II which the wave functions extend over the three spines of the helix. Although the two vibrons are trapped close to each other, there are not confined in the same spine so that the pair is allowed to explore the entire 3D nature of the helix. By contrast, at biological temperature, the dressing effect strongly reduces the vibrational exchanges between different spines so that only intraspine hopping constants remain. The pairs are thus confined in a given spine and exhibits the same features as the bound states described within the one-dimensional model (see Figs. 4). 
In a marked contrast with the previous situation, the occurrence of the third bound states is strictly due to the 3D nature of the helix. More precisely, as shown in Sec. II, this 3D nature manifests itself by the introduction of    
a generalized site dependent small polaron binding energy $E_{B}(n)$.  Although $E_{B}(n)$ takes significant values for $n=0$ and $n=3$, it does not strictly vanish for $n=1$. Note that $E_{B}(n) \approx 0$ for $n=2$ and $n>3$. As a consequence, Eq.(\ref{eq:HEFF}) clearly shows that this non vanishing value contributes to an attractive interaction between two vibrons located onto two different spines. However, the formation of TVBS-III results from the competition between the interspine hopping constants, which tends to break the pair, and the previous attractive interaction which allows for the trapping of the pair. At low temperature, the interspine hopping constants are strong enough to prevent the occurrence of the third bound states. By contrast, at biological temperature, the dressing effect strongly reduces the vibrational exchanges between different spines so that the attractive interaction mediated by $E_{B}(1)$ is sufficient to create the trapping of the two vibrons onto two different spines.  Note that strictly speaking the key parameter for the occurrence of the TVBS-III is $\hat{B}(1)$, which, in addition to its strong dependence on $E_{B}(1)$, exhibits a weak contribution $B(1)$ proportional to the intramolecular anharmonicity (see Eq.(\ref{eq:parameter2})). Due to the sign of the hopping constant $J(1)$, $B(1)$ is negative and tends to produce a repulsive interaction between the two vibrons. As a consequence, the binding energy of TVBS-III slightly decreases with the anharmonicity as shown in Figs. 5.  

\begin{figure}
\includegraphics{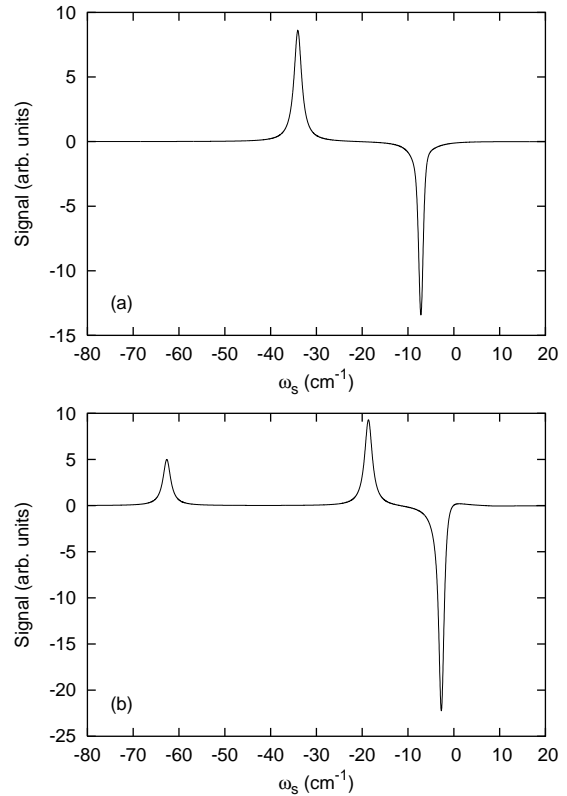}
\caption{Pump-probe spectrum of the amide-I band for (a) $T=310$ K, $\chi_{1}=25$ pN, $\chi_{3}=50$ pN, $A=8$ cm$^{-1}$ and (b) $T=310$ K, $\chi_{1}=37.5$ pN, $\chi_{3}=75$ pN, $A=8$ cm$^{-1}$.}
\label{fig:spectr}
\end{figure}

At this step, let us mention that this theoretical work clearly suggests the occurrence of two-vibron bound states for the amide-I vibrations in $\alpha$-helices. In a recent experiment, P. Hamm and co-workers \cite{kn:hamm1,kn:hamm2} have performed the infrared pump-probe spectroscopy of the N-H stretch vibrations in a stable $\alpha$-helix, the poly-$\gamma$-benzyl-L-glutamate (PBLG). They observed two excited-state absorption bands which were interpreted in terms of the occurrence of two kind of two-vibron bound states. In that context, the knowledge of both the single-vibron and the two-vibron dynamics allows us to simulate this pump-probe spectrum but for amide-I vibrations. 
During a pump-probe experiment, the helix interacts with two infrared short pulses, i.e. the pump and the probe pulse \cite{kn:hamm1,kn:hamm2,kn:hamm3,kn:hamm4,kn:hamm5,kn:hamm6}. The pump pulse excites the helix from its vibrational ground state to the first excited state. In other words, the pump creates a zero wave vector single vibron state. By contrast, the probe pulse measures the change of absorption as a result of that excitation. As a consequence, the pump-probe spectrum exhibits three contributions. The first contribution, called the Bleach, corresponds to the loss of absorption due to the depletion of the ground state whereas the second contribution characterizes the stimulated emission from the first excited state. These two processes lead to a negative peak in the pump-probe signal. The third contribution, which accounts for the creation of zero wave vector two-vibron states from the excited single-vibron state, yields a positive peak in the spectrum. From the knowledge of both the single-vibron \cite{kn:falvo} and  two-vibron states, it is straightforward to calculate the matrix element of the dipole involved in the three transitions characterizing the spectrum \cite{kn:hamm5}. To proceed, both the pump and the probe pulses are assumed to be polarized along the axis of the helix and an artificial width of 1 cm$^{-1}$ has been introduced to plot the spectrum. 

Fig. 7a shows the pump-probe bare spectrum at $T=310$ K and with the parameter values as in Fig. 3b. The spectrum, which is centered onto $2\hat{\omega_{0}}$, exhibits a single negative peak located at -7.2 cm$^{-1}$ and a single positive peak located at -34.0 cm$^{-1}$. In agreement with the result displayed in Fig. 3b, the positive peak is the signature of the zero wave vector bound state TVBS-I supported by the helix. Note that the frequency difference between the two peaks is about the binding energy of TVBS-I. 
For a stronger vibron-phonon coupling, i.e. for parameter values as in Fig. 5b, the pump-probe spectrum shows three peaks. A negative peak is located at -2.7 cm$^{-1}$ whereas two positive peaks occur at -18.6 cm$^{-1}$ and -62.6 cm$^{-1}$, respectively. As shown in Fig. 5b, the two positive peaks characterize the two bound states TVBS-I and TVBS-II. The intensity of the peak connected to TVBS-II is about two times greater than the intensity of the peak associated to TVBS-I. Note that the binding energy of TVBS-III is too weak so that the corresponding peak is embedded in the bleach. 

As illustrated in Figs. 7, a pump-probe experiment provides a direct evaluation of the strength of the different nonlinear sources favoring the occurrence of bound states through the measurement of both the number of positive peaks and the values of the corresponding binding energy. Nevertheless, let us mention that relaxation mechanism have been neglected in the present analysis so that the width of the different peaks has been disregarded. For amide-I vibration, we expect a competition between the binding energy of the bound states and the width of the corresponding peaks which will result in a difficulty to separate the different peaks. Note that it was not the case for N-H modes in PGLB because although the width of the different peaks was about 50 cm$^{-1}$, the binding energies of both TVBS-I and TVBS-II were about 275 and 120 cm$^{-1}$, respectively. These large values originate in the strong anharmonicity of each N-H stretching mode ($A=60$ cm$^{-1}$) as well as in the strong vibron-phonon coupling ($E_B(0)\approx 84$ cm$^{-1}$).

To conclude this discussion, let us address the fundamental question of the breather-like behavior of the two-vibron bound states. Indeed, the formation of discrete breathers in nonlinear classic lattices 
has been the subject of intense theoretical research during the last decade (for a recent review, see for instance Refs. \cite{kn:scott,kn:aubry,kn:flach,kn:mackay}). Discrete breathers, which correspond to
highly localized vibrational excitations in anharmonic lattices, do not require integrability for their existence and stability. They are not restricted to one-dimensional lattices and it has been suggested that they should correspond to quite general and robust time periodic solutions. Recently, the quantum nature of discrete breathers has been investigated and it has been suggested that bound states involving several quanta are good candidates for the quantum counterpart of discrete breathers \cite{kn:scott,kn:aubry,kn:flach,kn:mackay,kn:fleurov,kn:eilbeck,kn:dorignac,kn:proville}.

\begin{figure}
\includegraphics{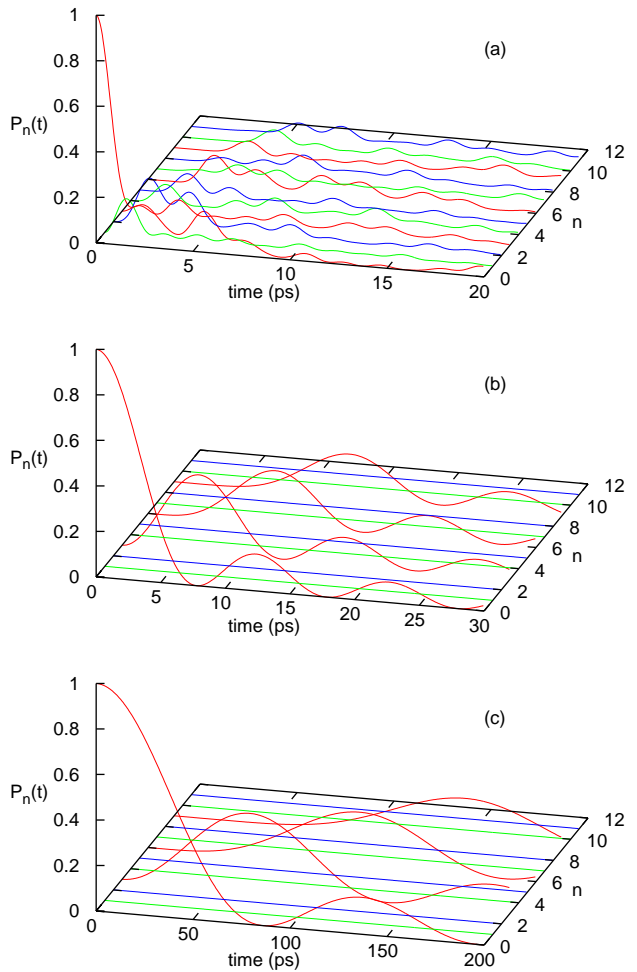}
\caption{Evolution of the probability $P_{n}(t)$ to observe the vibron pair on the site $n$ at time $t$. The pair, which has been created on the site $n=0$ at time $t=0$, is restricted to the TVBS-I band (see the text).(a) $T=5$ K, $\chi_{1}=25$ pN, $\chi_{3}=50$ pN, $A=8$ cm$^{-1}$, (b) $T=310$ K, $\chi_{1}=25$ pN, $\chi_{3}=50$ pN, $A=8$ cm$^{-1}$, (c) $T=310$ K, $\chi_{1}=37.5$ pN, $\chi_{3}=75$ pN, $A=8$ cm$^{-1}$.}
\end{figure}

To illustrate the concept of quantum breather in $\alpha$-helices, we restrict our attention to TVBS-I which are located below the TVFS continuum over the entire first Brillouin zone. As shown previously, these states correspond to the trapping of two vibrons onto the same amide-I mode and described two quanta which behave as a single particle. Note that, at this step, the breather-like behavior manifests itself in the localization of the separating distance between the two vibrons which form the pair. To characterize the motion of the pair, we use the procedure introduced by Proville \cite{kn:proville} which consists in defining the time dependent Wannier wave function as
\begin{equation}
\phi_{n}(t) = \frac{1}{\sqrt{N}} \sum_{K} e^{i(Kn-\omega_{K\lambda}t)} 
\end{equation}

When $\lambda$ is restricted to the TVBS-I band, $\phi_{n}(t)$ can be viewed as the wave function associated to the two trapped vibrons viewed as a single particle.
To characterize the motion of that particle, we introduce a coherent lifetime, proportional to the invert of the TVBS-I bandwidth, which measures the time spent by the pair on a given site. A zero bandwidth produces an infinite coherent lifetime for the pair which exhibits a breather-like behavior, i.e. the two vibrons
are trapped around the site where the pair has been created. By contrast, a finite bandwidth leads to the delocalization of the pair which moves along the lattice. 

By assuming that the pair is created onto the site $n=0$ at the initial time $t=0$, Figs. 8 show the behavior of the probability $P_n(t)=\mid \phi_n(t) \mid^{2}$ to observe the pair onto the site $n$ at time $t$. The three situations corresponding to the parameters used in Fig. 1b, 3b and 5b are illustrated in Fig. 8a, 8b and 8c, respectively. 
At low temperature, Fig. 8a clearly shows the 3D nature of the TVBS-I since the pair tends to rapidly delocalize over the three spines of the helix. Indeed, after a coherent lifetime of about 1.5 ps, 80 $\%$ of the initial population have left the excited site $n=0$ and have been tranfered to sites belonging to the other spines. Note that the population of the excited site shows damped oscillations due to the strong anistropy which discrimites between inter and intra spine vibrational exchanges. When the temperature reaches $T=310$ K, the confinement of the pair takes place as illustrated in Fig. 8b so that the pair propagates along the spine where it has been created. Nevertheless, the finite value of the TVBS-I bandwidth prevents the localization of the pair which leaves the excited site after a coherent lifetime of about 7.8 ps. However, for a stronger vibron-phonon coupling, the TVBS-I bandwidth is drastically reduced so that the breather-like behavior of the pair manifests itself by an enhancement of its local character (Fig. 8c). As previously, the pair is confined in the spine where it has been created but its coherent lifetime has been drastically increases and is about 86 ps. 
 
\section{conclusion}

In the present paper, the two-vibron dynamics associated to amide-I modes in an $\alpha$-helix has been described. The 3D nature of the helix was modeled by three spines of hydrogen-bonded peptide units linked via covalent bonds. 
To remove the intramolecular anharmonicity of each amide-I mode and to renormalize the strong coupling between the vibrons and the phonons associated to the external motion of the residues, two unitary transformations have been applied to finally obtain the dressed anharmonic vibrons point of view. 
In a perfect agreement with recent calculations performed within a one-dimensional model of helix, it has been shown that the two-vibron energy spectrum supports both a TVFS continuum and two kinds of bound states, called TVBS-I and TVBS-II, connected to the trapping of two vibrons onto the same amide-I mode and onto two nearest neighbor amide-I modes belonging to the same spine, respectively. 
In addition, it has been shown that the interplay between the temperature and the 3D nature of the helix strongly affects the characteristics of the two-vibron states. This feature originates in the singular behavior of the single vibron dynamics which displays a transition between two regimes. At low temperature, the single polaron behaves as an undressed vibron delocalized between the different spines whereas at biological temperature, the dressing effect strongly reduces the vibrational exchanges between different spines so that the polaron is confined in a single spine. The consequence is twofold. First, the nature of both TVBS-I and TVBS-II is strongly modified. Then, a third bound state called TVBS-III occurs. At low temperature, non vanishing interspine hopping constants yield a three dimensional nature of both TVBS-I and TVBS-II which the wave functions extend over the three spines of the helix whereas, at biological temperature, the pairs are confined in a given spine and exhibits the same features as the bound states described within a 1D model.
The occurrence of the TVBS-III results from the competition between the interspine hopping constants and the site dependent small polaron binding energy which favors an attractive interaction between vibrons located onto different spines. If at low temperature the hopping constants are strong enough to prevent the occurrence of TVBS-III, it is no longer the case at biological. As a result the attractive interaction becomes the dominant mechanism leading to the trapping of the two vibrons onto two different spines.  
Finally, the breather-like nature of the TVBS-I has been analyzed by studying the time evolution of a vibron pair initially created on a given site. Due to the 3D nature of TVBS-I, the pair rapidly delocalizes over the three spines at low temperature which results in a coherent lifetime of about a few picoseconds. By contrast, at biological temperature, the pair is confined in the spine where it has been created. Its coherent lifetime increases with the  strength of the vibron-phonon coupling and it has been shown that a strong coupling can localize the pair during a time of about 86 ps.


\begin{thebibliography}{99}
\bibitem{kn:aubry} S. Aubry, Physica \textbf{D103}, 201 (1997).
\bibitem{kn:flach} S. Flach and C.R. Willis, Phys. Rep. \textbf{295}, 181 (1998). 
\bibitem{kn:mackay} R.S. MacKay, Physica \textbf{A288}, 174 (2000).
\bibitem{kn:fleurov} V. Fleurov, Chaos \textbf{13}, 676 (2003).
\bibitem{kn:eilbeck} J.C. Eilbeck, Some exact results for quantum lattice problems, in \emph{Proceedings of the Third Conference Localization and Energy Transfert in Nonlinear Systems}, Edited by L. Vasquez, R.S. Mackay and M. P. Zorzano (World Scientific, Singapore, 2003). 
\bibitem{kn:kimball} J. C. Kimball, C. Y. Fong, and Y. R. Shen, Phys. Rev. \textbf{B23}, 4946 (1981).
\bibitem{kn:bogani} F. Bogani, G. Cardini, V. Schettino, and P. L. Tasselli ,  Phys. Rev. \textbf{B42}, 2307 (1990).
\bibitem{kn:scott} A. C. Scott, J. C. Eilbeck, and H. Gilhoj, Physica \textbf{D78}, 194 (1994).
\bibitem{kn:davydov}  A. S. Davydov and N. I. Kisluka, Phys. Status Solidi \textbf{59}, 465 ; Zh. Eksp. Teor. Fiz \textbf{71}, 1090 (1976) [Sov. Phys. JETP \textbf{44}, 571 (1976)].
\bibitem{kn:davydov1} A.S. Davydov, \emph{Soliton in Molecular Systems} (D. Reidel, Dordrecht, 1985).
\bibitem{kn:scott1} A.C. Scott, Phys. Rep. \textbf{217}, 1 (1992).
\bibitem{kn:chris} P. L. Christiansen and A. C. Scott , \emph{Davydov's Soliton Revisited},(Plenum, New York, 1990).
\bibitem{kn:brown1}D.W. Brown and Z. Ivic, Phys. Rev. \textbf{B40}, 9876 (1989). 
\bibitem{kn:brown2}D.W. Brown, K. Lindenberg, and X. Wang, in \emph{Davydov's Soliton Revisited}, edited by P. L. Christiansen and A. C. Scott (Plenum, New York, 1990), .
\bibitem{kn:ivic1}Z. Ivic, D. Kapor, M. Skrinjar, and Z. Popovic, Phys. Rev. \textbf{B48}, 3721 (1993).
\bibitem{kn:ivic2}Z. Ivic, D. Kostic, Z. Przulj, and D. Kapor, J. Phys. Condens. Matter \textbf{9}, 413 (1997).
\bibitem{kn:ivic3}J. Tekic, Z. Ivic, S. Zekovic, and Z. Przulj, Phys. Rev. \textbf{E60}, 821 (1999).
\bibitem{kn:falvo} C. Falvo and V. Pouthier, J. Chem. Phys., preceding paper submitted (2005).
\bibitem{kn:pouthier1} V. Pouthier, Phys. Rev. \textbf{E68}, 021909 (2003).
\bibitem{kn:pouthier2} V. Pouthier and C. Falvo, Phys. Rev. \textbf{E69}, 041906 (2004).
\bibitem{kn:hamm1} J. Edler, R. Pfister, V. Pouthier, C. Falvo and P. Hamm, Phys. Rev. Lett. \textbf{93}, 106405 (2004).
\bibitem{kn:hamm2} J. Edler, V. Pouthier, C. Falvo, R. Pfister and P. Hamm in \emph{Ultrafast Phenomena XIV}, edited by T. Kobayashi, T. Okada, T. Kobayashi, K. Nelson, S. De Silvesti, 
Springer Series in Chemical Physics, Vol. \textbf{79} (Springer, Berlin, 2005). 
\bibitem{kn:hennig} D. Hennig, Phys. Rev. \textbf{B65}, 174302 (2002).
\bibitem{kn:chris2} P. L. Christiansen, A.V. Zolotaryuk, and A.V. Savin, Phys. Rev. \textbf{E56}, 877 (1997).
\bibitem{kn:lang} I. G. Lang and Yu. A. Firsov, Sov. Phys. JETP \textbf{16} , 1293 (1962).
\bibitem{kn:pouthier01} V. Pouthier, and C. Girardet, Phys. Rev. \textbf{B65}, 035414 (2002).
\bibitem{kn:pouthier02} V. Pouthier, J. Chem. Phys. \textbf{118}, 3736 (2003).
\bibitem{kn:pouthier03} V. Pouthier, J. Chem. Phys. \textbf{118}, 9364 (2003).
\bibitem{kn:scott2} A.C. Scott, Phys. Rev. \textbf{A26}, 578 (1982).
\bibitem{kn:hamm3} P. Hamm, M. Lim, and R. M. Hochstrasser J. Phys. Chem. \textbf{B102}, 6123 (1998).
\bibitem{kn:hamm4} S. Woutersen and P. Hamm, J. Phys.: Condens. Matter \textbf{14}, R1035 (2002).
\bibitem{kn:muk} A.M. Moran, S.M. Park, J. Dreyer, and S. Mukamel, J. Chem. Phys. \textbf{118}, 3651 (2003)   
\bibitem{kn:hamm5} J. Edler and P. Hamm, J. Chem. Phys. \textbf{117}, 2415 (2002)   
\bibitem{kn:hamm6} J. Edler and P. Hamm, J. Chem. Phys. \textbf{119}, 2709 (2003).
\bibitem{kn:dorignac} J. Dorignac, J.C. Eilbeck, M. Salerno and A.C. Scott, Phys. Rev. Lett. \textbf{93},025504 (2004). 
\bibitem{kn:proville} L. Proville, Phys. Rev. \textbf{B71}, 1043306 (2005).
\end{thebibliography}
\end{document}